\documentclass[12pt,fleqn]{article}
\usepackage{xiiiemc}
\usepackage{natbib}
\usepackage{fancyhdr}
\usepackage{color}
\usepackage{wallpaper} 
\usepackage{titlesec}   %% Define space between paragraph e section
\usepackage{float} 	%% Use to fix Figure or Table: ex: \begin{table}[H]
\usepackage[usenames,dvipsnames]{xcolor}
\usepackage{hyperref} 	%% Use to fix Figure or Table: ex: \begin{table}[H]
\hypersetup{
	pdfcreator={LaTeX with abnTeX2},
	pdfkeywords={abnt}{latex}{abntex}{USPSC}{trabalho acadêmico}, 
	colorlinks=true,       		% false: boxed links; true: colored links
	linkcolor=blue,          	% color of internal links
	citecolor=blue,        		% color of links to bibliography
	filecolor=magenta,      		% color of file links
	urlcolor=blue,
	allbordercolors=black,
	bookmarksdepth=4
}
\usepackage{tocloft}
\titleformat{\section}
  {\normalfont\bfseries}{\thesection.}{0.5em}{}
 
\renewcommand\thesection{\arabic{section}}

\let\OLDthebibliography\thebibliography
\renewcommand\thebibliography[1]{\OLDthebibliography{#1} \setlength{\parskip}{0pt}\setlength{\itemsep}{0pt plus 0.3ex}}

\title{TEXT MINING DESCRIPTIONS OF DREAMS: AESTHETIC AND CLINICAL EFFORTS}

\author
    {\rm \begin{tabular}{l} 
    \textbf{Renato Fabbri}$^{1}$ - {\textnormal renato.fabbri@gmail.com}\\%
    \textbf{Fabiane M. Borges}$^{2}$ - {\textnormal catadores@gmail.com}\\
    {\fontsize{11}{0}\selectfont $^{1}$University of São Paulo, Institute of Mathematical and Computer Sciences - São Carlos, SP, Brazil}\vspace*{-0.05cm} \\
    {\fontsize{11}{0}\selectfont $^{2}$Federal University of Rio de Janeiro, School of Fine Arts - Rio de Janeiro, RJ, Brazil}\vspace*{-0.05cm}\\
  \end{tabular}}
\fancypagestyle{firspagetstyle}
{
	\lhead{}
	\fancyhead[C]{%
		\includegraphics[width=0.9\linewidth]{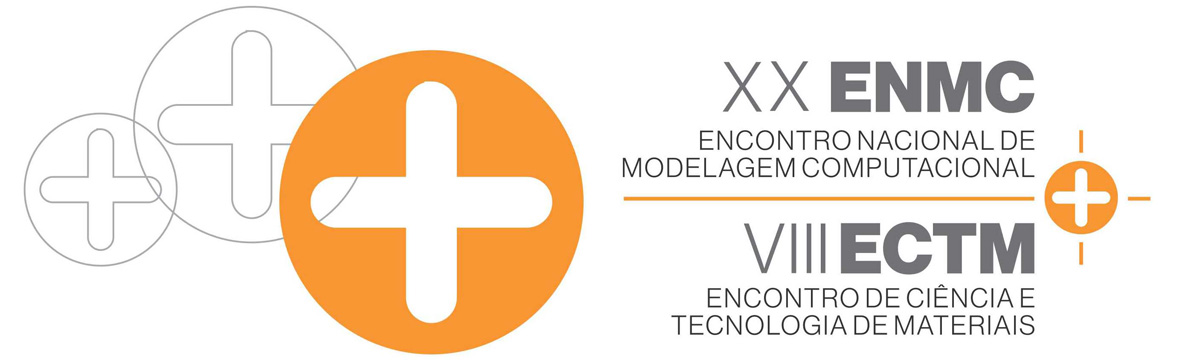}\\%
		{\scriptsize \fontfamily{phv}\fontseries{b}\selectfont \color[rgb]{0.45,0.45,0.45}
		16 a 19 de Outubro de 2017\\
		Instituto Politécnico - Universidade do Estado de Rio de Janeiro\\
		Nova Friburgo - RJ\\
	    }
	}
	\renewcommand{\headrulewidth}{0.0pt}
	\fancyfoot[C]{\footnotesize \parbox{15cm} {\centering  \fontsize{7.5}{0}\selectfont \it Anais do XX ENMC – Encontro Nacional de Modelagem Computacional e VIII ECTM – Encontro de Ciências e Tecnologia de Materiais,  Nova Friburgo, RJ – 16 a 19 Outubro 2017}} % \ttfamil
	\rhead{}
}

\begin{document}
\maketitle

\thispagestyle{firspagetstyle}

\fancyhead[L]{\footnotesize{\fontsize{7.5}{0}\selectfont \it XX ENMC e VIII ECTM\\
	16 a 19 de Outubro de 2017\\
	Instituto Politécnico Universidade do Estado do Rio de Janeiro – Nova Friburgo - RJ\\}}
\renewcommand{\headrulewidth}{0.0pt}
\fancyfoot[C]{\footnotesize \parbox{15cm} {\centering  \fontsize{7.5}{0}\selectfont \it Anais do XX ENMC – Encontro Nacional de Modelagem Computacional e VIII ECTM – Encontro de Ciências e Tecnologia de Materiais,  Nova Friburgo, RJ – 16 a 19 Outubro 2017}} % \ttfamil
\rhead{}

\begin{abstract}
Dreams are highly valued in both Freudian psychoanalysis and less conservative clinical traditions.
Text mining enables the extraction of meaning from writings in powerful and unexpected ways.
In this work, we report methods, uses and results obtained by mining descriptions of dreams.
The texts were collected as part of a course in Schizoanalysis (Clinical Psychology) from dozens of participants.
They were subsequently mined using various techniques for the achievement of poems and summaries,
which were then used in clinical sessions by means of music and declamation.
The results were found aesthetically appealing and effective to engage the audience.
The expansion of the corpus, mining methods and strategies for using the derivatives for art
and therapy are considered for future work.
\end{abstract}

\keywords{\em{Text mining, Dreams, Schizoanalysis, Poetry, Art}}

\pagestyle{fancy}

\section{INTRODUCTION}
Although dreams are described in texts that range from ancient sacred~\citep{bible,boas,kopenawa}
to psychiatric~\citep{dreamMed}, there is no consensus of what dreams are.
% * <renato.fabbri@gmail.com> 2017-08-22T10:30:03.356Z:
% 
% ae meu comentário aqui
% 
% ^.
We can exemplify the diverse theories with three simple cases~\citep{dreamsGen}:
\begin{itemize}
	\item dreams are often regarded by the dreamers as accessing spiritual realms and other realities.
	\item Many scientists regard dreams as by-products of the sleeping process:
		arbitrary interpretations given by the conscious mind to noisy signals without substantial meaning.
	\item Freudian and Jungian psychoanalytic traditions understand dreams as symbolic constructs output by the unconscious mind.
\end{itemize}
\noindent We can, however, state some facts about dreams that assert them as attractive for therapy and for art.
First, dreams are often very rich in impacting and symbolic images.
Second, they are told by the person who dreamed in a very attentive manner, as being very significant to the dreamer.
In fact, most of us should be able to recall a number of situations where someone (perhaps ourselves)
was describing a dream in a rapid, almost euphoric, succession of words.
Dreams are so effective in yielding artistic materials that surrealism is an aesthetic explicitly inspired by dreams and
symbolism is an example of artistic movement heavily influenced by dreams.

Text mining is data mining applied to textual data.
There are many models for the text mining pipeline, but
it can be summarized as: data collection and preparation,
pattern recognition, evaluation of the output and reporting~\citep{tmining}.
This work addresses text mining of descriptions of dreams
with aesthetic and therapeutic purposes.

Section~\ref{sec:matMet} describes the corpus and methods.
Section~\ref{sec:res} is dedicated to presentation and discussion of results.
Section~\ref{sec:conc} holds conclusions and further work considerations.

\section{MATERIALS AND METHODS}\label{sec:matMet}
\subsection{Corpus}
The description of dreams we used are all in Brazilian Portuguese,
collected as part of clinical practices in the year of 2015.
Participants described the dreams and sent them to the second author of this paper, who is a psychologist.
Thereafter, another collection of dreams was gathered in the same way by the second author and collaborators,
with the purpose of expanding the analysis and synthesis of texts performed with the previous
corpus.
It is a larger corpus, also in Brazilian Portuguese.
Interestingly, both corpus contain descriptions of dreams by women only.
Their scales are summarized in Table~\ref{tab:dreams} in terms of numbers of characters, tokens,
and paragraphs.

\begin{table}[H] % !htbp 
	\caption{Corpus files and elementary countings.
	The number of dreams is about the same as the number of paragraphs.
	The date column corresponds to  the month when the last dream was collected.}\label{tab:dreams}
\vspace{12pt}
\centering{}
	\begin{tabular}{  c || c | c | c | c | c }
		\textbf{File}           & \textbf{Characters} & \textbf{Tokens} & \textbf{Sentences} & \textbf{Paragraphs} & \textbf{Date} \\\hline
		\texttt{corpora.txt}  & 9456 & 1693 & 104 & 30 & Mar/2015 \\
		\texttt{corpora2.txt}  & 71514 & 14691 & 435 & 156 & Nov/2015 \\
\end{tabular}
\end{table}

\subsection{Analysis and derivation methods}
The texts were analyzed to support the extraction of meaning from the dreams
and for the creation of artistic texts.
We strived to keep the methods very simple in order to avoid puzzling the involved parties.
Three lists of tokens were considered:
\begin{itemize}
	\item punctuations !"\#\$\%\&'()*+,-./:;<=>?@[]\textbackslash\textasciicircum\_`\{\}|\textasciitilde. Obtained through the command \\ \texttt{string.punctuation} of Python's string (standard) library.
	\item Portuguese \emph{stopwords}\footnote{The exact definition and list of stopwords are not consensual.
		Anyway, one can regard them as words with lesser meaning and which are very frequent, such as conjunctions and prepositions.}
		obtained through NLTK~\citep{nltk} by the command \\ \texttt{nltk.corpus.stopwords.words("portuguese")}.
	\item Tokens in the texts which were not punctuations nor stopwords.
		These were regarded as the most meaningful words in the corpus.
\end{itemize}

This selection of most meaningful words was used as the core material for the achievement
of more interesting constructions for
art and clinical psychology  through filtering and ordering.
Most significantly, the ordering could be based on the alphabet, the size of tokens in number of letters,
or the count of incidences of the words, or any combination of these.
Filtering could be performed by restricting the vowels, consonants (e.g. fricatives), word size, frequency, or
collocations.

\section{RESULTS AND DISCUSSION}\label{sec:res}
The list of most meaningful words (described in the previous section) was filtered and ordered
in many ways to yield diverse sequences of interest.
After an inspection of the results, these criteria were selected to compose a final document:
\begin{itemize}
	\item Ordering by: incidence (most frequent words first),
alphabetic, size in characters, with and without repetitions.
These were considered the most raw sequences and used subsequently to derive other sequences
		with such variations of ordering and repetition.
	\item Words with only one vowel (repeated any number of times).
	\item Only words with fricatives or plosives or
some combination of them (e.g. plosives and 'm' and vowels 'a' and 'e').
	\item Words that start and end sentences.
	\item Collocations (pairs of words which are frequent together).
\end{itemize}

Such final document and other files are available online and exposed in Table~\ref{tab:files}.
An example of the derived texts is in Table~\ref{tab:comecoFinal} with a translation
from Portuguese to English.
These texts were used for aesthetic appreciation and also in a schizoanalysis group session
in 2014 (Casa Nuvem, Rio de Janeiro, RJ, Brazil).
In the same course, the artist Giuliano Obici used the texts to feed his program "Voices Simulacrum"~\citep{obici}: a machinic chorus of robotic voices with computers connected in network. The project integrated devices in the computer (sound, video and network cards) exploring the network as a distributed audiovisual instrument (conceptualized as a ``metamedia-instrument'')
to read the texts in a very performative manner.

The group was constituted by the participants who described the dreams and
the report of the episode is somewhat impressive:
the members had strong impressions, some of them cried and entered a quasi-shock state. 

\begin{table}[H] % !htbp 
	\caption{Files related to the text mining of dreams.
	All files are found in a public git repository dedicated to the developments presented in this article~\citep{repo}.}\label{tab:files}
\vspace{12pt}
\centering{}
	\begin{tabular}{  c | p{8cm} }
	\textbf{File}           & \textbf{Description} \\\hline
	\texttt{scripts/todos.py}  & Python script that makes the current analysis and renders the TXT and PDF files.     \\
	\texttt{corpora/corpora.txt}  & The first collection of descriptions of dreams. \\
		\texttt{corpora/corpora2.txt}  & The second (and larger) collection of descriptions of dreams.      \\
	\texttt{mineracaoDosSonhos/PLNSonhos.odt}  & A brief consideration of the text mining of dreams to which this article is dedicated.  \\
		\texttt{mineracaoDosSonhos/TUDO.pdf}  & A thorough exposition of all the (selected) texts derived from the descriptions of dreams.  \\
\end{tabular}
\end{table}

\newpage %

\vspace{12cm}

\begin{table}[H] % !htbp 
	\caption{Example of artistic text achieved from the descriptions of dreams.
	This text was obtained through picking only the first and last words of each sentence.
	As illustrated in this text, the unusual (and formally wrong) morphological and syntactic
	constructions were used for enhanced artistic expression
	and as clinical evidence of complex cognitive elaborations.}\label{tab:comecoFinal}
\vspace{2pt}
\centering{}
 \small
\begin{tabular}{  c | c }
	\textbf{Portuguese (original)}      &                                                \textbf{English (translation)} \\
Escorregava glandes              &                                                Slipping glands         \\
Numa assustavam                  &                                                At once, they scared    \\
Eu suada                         &                                                I sweated               \\
As cavalos                       &                                                The horses              \\
Não acabou                       &                                                It's not over           \\
Barras mim                       &                                                Bars me                 \\
Andei construtores               &                                                I walked builders       \\
Pessoas )                        &                                                People  )               \\
  & \\
Sonhei formei                    &                                                I dreamed I formed      \\
Estava menino                    &                                                It was boy              \\
Depois boa                       &                                                Then good               \\
Esse meu                         &                                                This mine               \\
Sonhos descendo                  &                                                Dreams coming down      \\
  & \\
O irmão                          &                                                The brother             \\
Meu punição                      &                                                My punishment           \\
Começa irmão                     &                                                Begins brother          \\
Meu ele                          &                                                My him                  \\
Meu demonstração                 &                                                My demonstration        \\
Depois ”                         &                                                After "                 \\
Eu parede                        &                                                I wall                  \\
Sinto dele                       &                                                I'm fell him            \\
A importência                    &                                                The ``importence''      \\
O buraco                         &                                                The hole                \\
  & \\
Acordei ofegante                 &                                                Woke up breathless    \\
Sensação NÃO                     &                                                Feeling NO              \\
Já rumo                          &                                                I'm on my way           \\
Estava perseguido                &                                                Was persecuted          \\
Quando percebeu                  &                                                When realized    \\
A tempo                          &                                                In time                 \\
Até porta                        &                                                Up to door              \\
O disso                          &                                                The this                    \\
Eu sobreviver                    &                                                I survive               \\
	Parecia ferramentas              & Seemed like tools                             \\
  & \\
	Três mim          & Three of me \\
	Lavo piano         & I wash piano \\
	Havia tirano & There was a tyrant \\
	Nele tudo & In him all \\
	Jogaram fogo & They set fire \\
	Pessoas presas &  People trapped \\
	Comecei ali & I started there \\
	Mas destruísse & But destroy \\
	Pessoas criança & Children people
\end{tabular}
\end{table}

\newpage %

% \newpage %
\section{CONCLUSIONS AND FUTURE WORK}\label{sec:conc}
We understand that the results are compelling for both art and clinical psychology.
Only the first corpus was used, which is smaller and made easier the selection of the resulting texts.
The methods applied are very simple, favoring the communication between the parties,
and are promptly deepened and expanded into more complex processes.
This work seems unique in the sense of using text mining of dreams for art and clinical psychology,
which, in our opinion, benefits the appreciation of it as a multidisciplinary and scientific contribution
in computer science, art and psychology.

In further efforts,
we might use for the corpus:
\begin{itemize}
	\item descriptions of dreams in the literature (e.g. from the mythology, traditional communities, etc); 
	\item other languages;
	\item an expansion of current corpus;
	\item dreams from specific groups, e.g. again gender related or of a specific age span, professional or educational background, etc.
\end{itemize}

\noindent About the text mining methods, we might:
\begin{itemize}
	\item use specific routines for classification (e.g. clusterization) of texts or their features;
	\item expand the methods of selection of words to better encompass meter (e.g. number of syllables);
	\item expand the methods of selection of words and phrases by their sonorities (e.g. by using sequences of vowels or consonants, mute consonants, paroxytones);
	\item use Wordnet~\citep{wordnet} in order to relate terms through semantic links (e.g. hypernymy, meronymy, synonymy).
\end{itemize}

The exploration of the results in therapeutic sessions and for the achievement of collections of
artistic texts should be kept as the core purposes.

\subsection*{\textit{Acknowledgements}}
The authors thank the volunteers who supplied the descriptions of dreams;
the subjects who attended to the schizoanalysis sessions;
the open source software developers, especially those who enabled this work by developing
the Python language and the NLTK.

% ------------------------------------------------------------------------

% ------------------------------------------------------------------------

%For papers written in Portuguese or Spanish.

%\begin{center}
%  TITLE IN ENGLISH
%\end{center}

%\def\abstractname{Abstract}%

%\begin{abstract}
%   Abstract in english
%\end{abstract}

%\keywords{\em{Keywords in english}}

\end{document}